\documentclass[aps,prl,10pt,twocolumn,amsmath,showpacs,amssymb,superscriptaddress,nofootinbib]{revtex4-1}

\usepackage{graphics,graphicx}
\usepackage{amsmath, amssymb}
\usepackage{multirow}
\usepackage{longtable}
\usepackage{color}
\usepackage{hyperref}

\usepackage[normalem]{ulem}  

\begin{document}

\title{The  thermal proton yield anomaly in Pb-Pb collisions at the LHC and its  resolution}
\date{\today}

\author{Anton Andronic}
\affiliation{Institut f\"ur Kernphysik, Universit\"at M\"{u}nster, D-48149 M\"{u}nster, Germany}
\author{Peter Braun-Munzinger}
\affiliation{Extreme Matter Institute EMMI, GSI-Helmholtzzentrum f\"ur Schwerionenforschung, D-64291 Darmstadt, Germany}
\affiliation{Physikalisches Institut der Universit\"at Heidelberg, D-69120 Heidelberg, Germany}
\affiliation{Institute of Particle Physics and Key Laboratory of Quark and Lepton Physics (MOE),
 Central China Normal University, Wuhan 430079, China
}
\author{Bengt Friman}
\affiliation{GSI-Helmholzzentrum f\"{u}r Schwerionenforschung, D-64291 Darmstadt, Germany}
\author{Pok Man Lo}
\affiliation{Institute of Theoretical Physics, University of Wroclaw,
PL-50204 Wroc\l aw, Poland}
\author{Krzysztof Redlich}
\affiliation{Institute of Theoretical Physics, University of Wroclaw,
PL-50204 Wroc\l aw, Poland}
\affiliation{Fakult\"at f\"ur Physik, Universit\"at Bielefeld, D-33615 Bielefeld, Germany}
\affiliation{Extreme Matter Institute EMMI, GSI-Helmholtzzentrum f\"ur Schwerionenforschung,
D-64291 Darmstadt, Germany}
\author{Johanna Stachel}
\affiliation{Physikalisches Institut der Universit\"at Heidelberg,
D-69120 Heidelberg, Germany}
\affiliation{Extreme Matter Institute EMMI, GSI-Helmholtzzentrum f\"ur Schwerionenforschung,
D-64291 Darmstadt, Germany}

\begin{abstract}
We propose a resolution of the  discrepancy 
between the proton yield predicted by 
the statistical hadronization approach and
data on hadron production in ultra-relativistic nuclear collisions
at the LHC. Applying the S-matrix formulation of statistical mechanics to include pion-nucleon interactions,
we reexamine their contribution to the proton yield,
taking resonance widths and the presence of nonresonant correlations into account. The effect of multi-pion-nucleon interactions is estimated using lattice QCD results on the baryon-charge susceptibility.
We show that a consistent implementation of these features
in the statistical hadronization model, leads to a reduction of the predicted proton yield,
which then quantitatively matches data of the ALICE collaboration for
Pb-Pb collisions at the LHC.

\end{abstract}

\maketitle

\section{Introduction}

\begin{figure*}[ht!]
\includegraphics[width=3.355in]{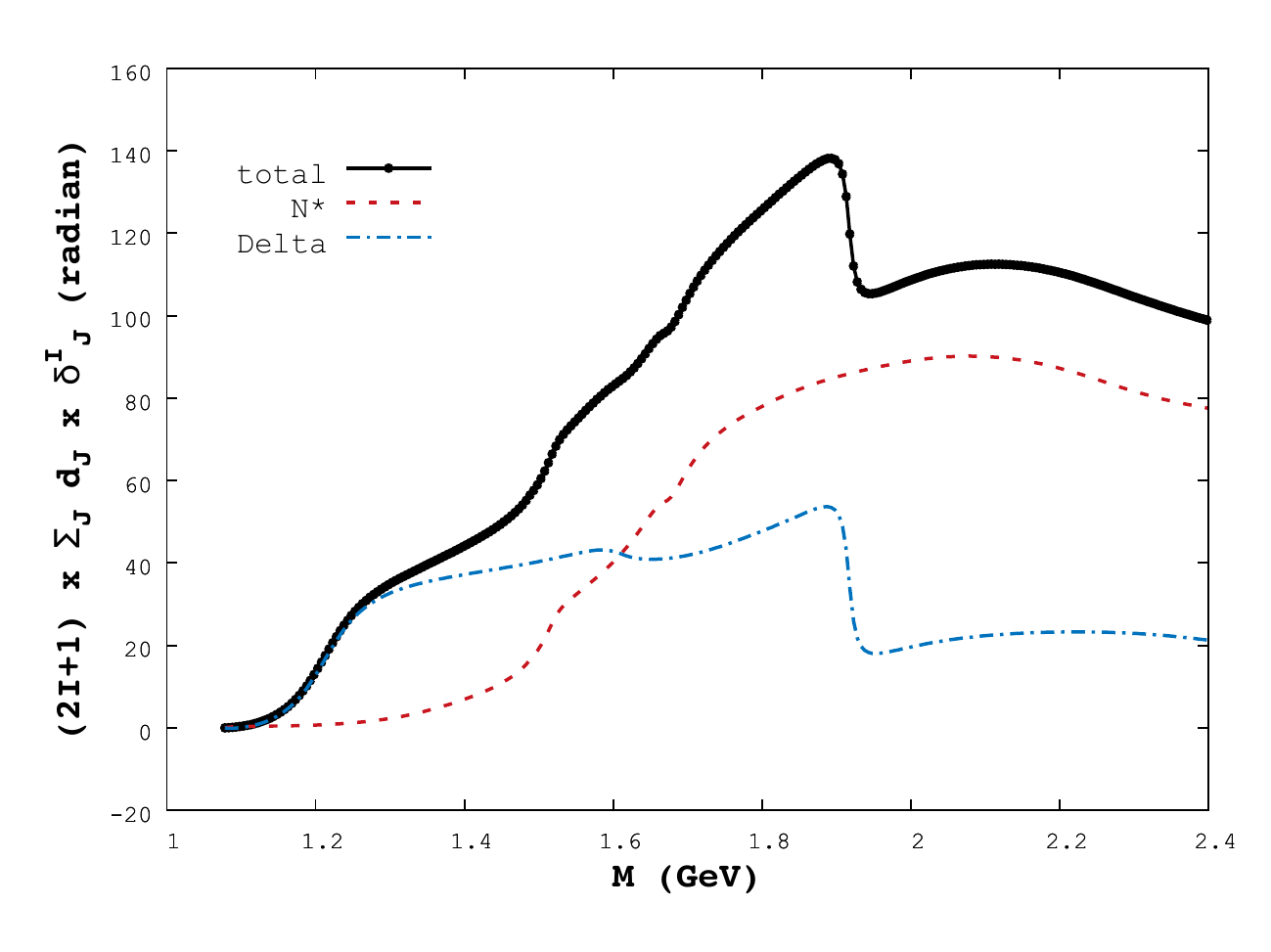}
\includegraphics[width=3.355in]{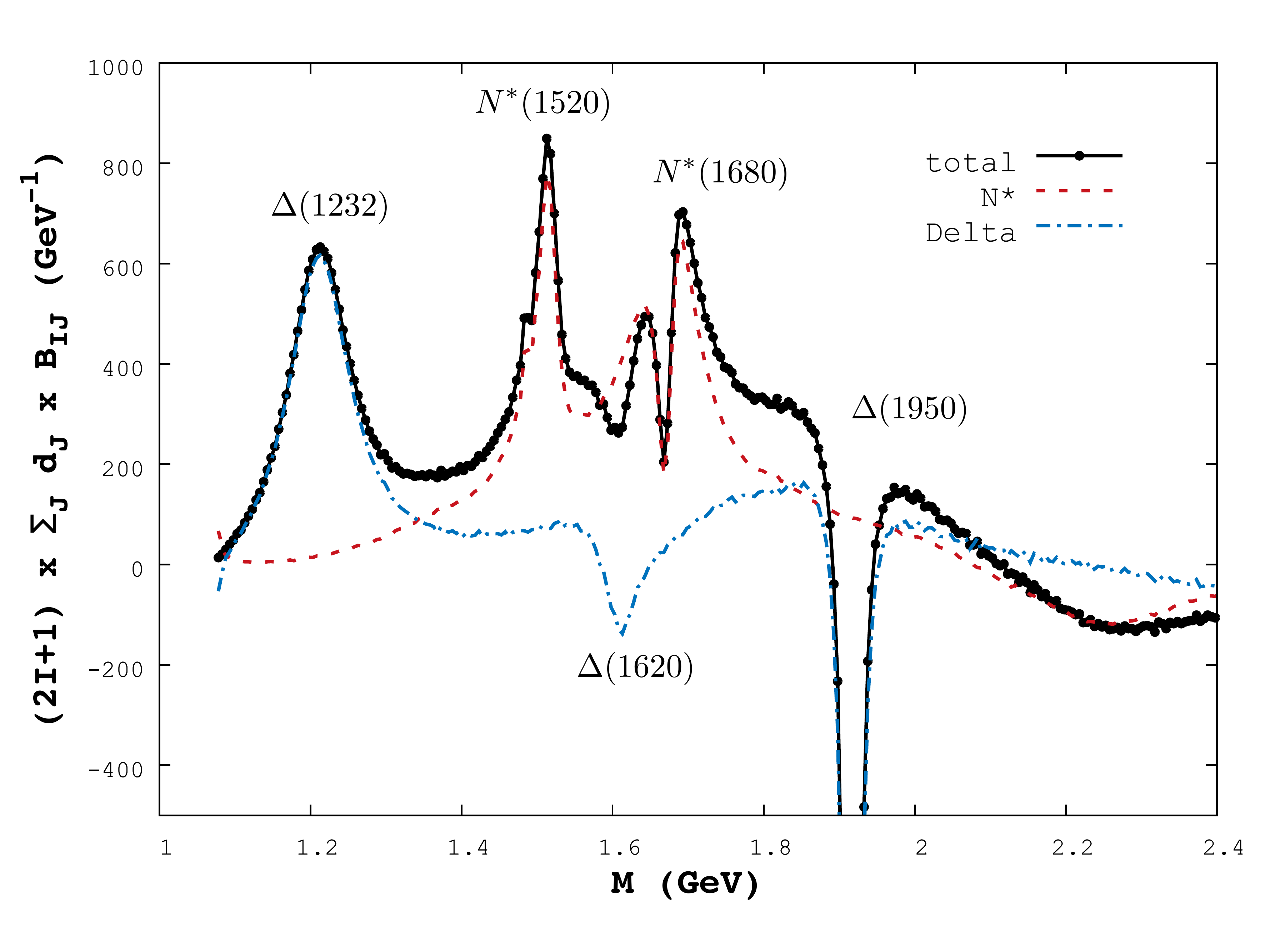}
\caption{(color online). On the left, we show the spin-isospin-weighed sums of the empirical $\pi N$ phase shifts in the $I=1/2$ ($N^*$) and $I=3/2$ ($\Delta$) channels, obtained from the GWU/SAID partial-wave analysis~\cite{Workman:2012hx}. The corresponding effective spectral functions, computed using Eq.~\eqref{eqn:bfunc}, are presented on the right. In this study, 15 partial waves in each isospin channel have been included in the effective spectral function. Structures associated with prominent resonances are indicated for reference. As discussed in the text, some resonances produce dips rather than peaks.}
\label{fig:fig1}
\end{figure*}
The thermal nature of particle production in high energy nucleus-nucleus  collisions is one of the important findings in the phenomenological analysis of experimental data \cite{Bnature,BraunMunzinger:2003zd,statmod,stat0,stat1,stat2,stat3,stat4,stat5}. Particle yields measured in heavy-ion collisions  in a very broad energy range closely follow  thermal equilibrium populations of a hadron resonance gas (HRG). At low temperatures, $T < 160$ MeV, the HRG  also reproduces the equation of state obtained from  Lattice QCD (LQCD) calculations \cite{hlqcd1,hlqcd2}.   Consequently, the statistical hadronization approach provides  a link between data on hadron production in ultra-relativistic nuclear collisions and the QCD partition function \cite{Bnature}.

The thermal origin  of particle production  in heavy ion collisions  is particularly transparent at LHC energies, where the chemical freeze-out is quantified by only two parameters, the temperature   and the volume   of the produced fireball.
The  temperature parameter of the statistical hadronization approach, which has been obtained with considerable precision by comparing to the yields of particles measured by the ALICE collaboration, coincides within errors with the pseudocritical temperature of the chiral crossover transition quantified in LQCD \cite{Bnature,lhc1,Aoki:2006we,Bazavov:2014noa,Steinbrecher:2018phh}.

There is an impressive overall agreement over 9 orders of magnitude between particle multiplicities measured at the LHC and the statistical hadronization ana\-lysis. The only conspicuous difference between data and calculations is observed for the proton and antiproton yields, where a deviation of 2.7 standard deviations is observed~\cite{lhc1}. The statistical hadronization model predicts about 25\% more protons and antiprotons than measured by the ALICE collaboration in central Pb--Pb collisions at the LHC. This constitutes the much debated ``proton-yield anomaly'' in heavy-ion collisions at the LHC.

The above  difference was argued
to  be possibly connected to the annihilation of baryons in the hadronic phase after chemical freeze-out \cite{stock}. However, deviations between the HRG model  and  LQCD results were also observed on the level of the second-order electric-charge and baryon-number susceptibility $\chi_{BQ}$,  \cite{chiBQ} where contributions of protons and antiprotons  and resonances  play an important  role. This led to the conjecture that the observed ``proton-yield anomaly'' could be due to the cursory treatment of interactions  in  the statistical operator of the hadron resonance gas model.

Generally, the proton yields predicted by statistical hadronization can be separated into two parts: a purely thermal yield from uncorrelated nucleons, which depends only on the freeze-out parameters, i.e., the freeze-out temperature $T_f$ and the freeze-out chemical potentials, and the contribution from multiparticle interactions involving nucleons. The latter include baryon resonances, as well as nonresonant meson-baryon interactions. 

In the statistical approach to particle production,   
the resonance contribution is commonly estimated using the HRG model.
This model assumes that resonance formation dominates the interactions of the confined phase,
and treats the resonances as an ideal gas, with masses as reported by the Particle Data Group~\cite{pdg}.
However, many of the simplifying assumptions implicit in the HRG model are not necessarily consistent with hadron scattering data.
In particular, for an accurate determination of 
interaction effects, resonance widths and the presence of many nonresonant contributions cannot be neglected.
Thus, to describe the high-precision hadron yield data from ALICE at LHC, a more refined approach is required to properly account for  the interaction contributions to particle multiplicities. It is the purpose of this letter to employ a consistent theoretical framework to reliably describe  resonant and nonresonant pion-nucleon interactions and their contributions to the proton yield.

The S-matrix formalism~\cite{dmb, Prakash, weinhold, rho, smat}
is a systematic framework for incorporating interactions into
the description of the thermal properties of a dilute medium.
In this scheme, two-body interactions 
are, via the scattering phase shifts, included in the leading term of the S-matrix expansion of the grand canonical potential. The resulting interacting density of states is then folded into an integral over thermodynamic distribution functions, which, in turn, yields the interaction contribution to a particular thermodynamic observable. The calculation of higher-order terms in the expansion is, as a rule, much more involved, since the required three- and higher-body S-matrices~\cite{dmb,smat} are in general not readily available.

Recently, the S-matrix approach has been used
to study the baryon-charge susceptibility $\chi_{BQ}$ in 
a thermal medium~\cite{chiBQ}. It was demonstrated that the  
implementation of the empirical pion-nucleon phase shifts
is crucial for the proper interpretation of the 
LQCD result. Also in the analysis of observables involving nucleons in ultra-relativistic nucleus-nucleus collisions, 
a careful handling of $\pi N$ interactions is mandatory, 
since the pion is the most abundant particle at freeze-out and the corresponding cross section 
is large. It is therefore natural to ask whether a systematic treatment of pion-nucleon interactions
in the partition function can  resolve the ``proton-yield anomaly'' introduced above.  

In this letter, we show that the answer to the above question is indeed affirmative and demonstrate that the S-matrix treatment of pion-nucleon and multi-pion-nucleon interactions removes the discrepancy between experimental data and theoretical predictions within the statistical hadronization approach. The contribution of the multi-particle channels is assessed by employing a connection between the proton yield and the baryon-charge susceptibility. The resulting proton multiplicity is in good agreement with the ALICE data.

\section{S-matrix treatment of the $\pi N$ system.}

To determine  
the effect of $\pi N$ interactions on the proton and antiproton abundances,
we apply the leading term in the S-matrix expansion for the grand canonical potential. 
The merit of this approach, when used in conjunction with 
empirical phase shifts, deduced from scattering experiments,
is that it offers a consistent, model-independent way to incorporate both resonant and nonresonant interactions between hadrons.
This is particularly important for studying the $\pi N$ system,
due to significant nonresonant contributions to the S-matrix and the occurrence of several prominent broad resonances.
As we shall demonstrate, a proper treatment of these effects is crucial for an accurate determination of the proton yield.

A detailed S-matrix analysis of the $\pi N$ system has been presented in Ref.~\cite{chiBQ}.
The thermal yield of a $\pi N$ interaction channel with spin $J$ and isospin $I$ is given by~\cite{weinhold, smat}
\begin{align}
\label{eqn:res}
  \begin{split}
    \langle R_{J,I} \rangle &= d_{J} \int_{m_{th}}^\infty d M  \, \frac{1}{2 \pi} \, B_{J,I}(M)  \\
    &\times \int \frac{d^3 p}{(2 \pi)^3}\,\frac{1}{e^{(\sqrt{p^2+M^2}-\mu)/T}+1},
  \end{split}
\end{align}
\noindent where
\begin{align}
  \label{eqn:bfunc}
  B_{J,I}(M) =  2 \, \frac{d \delta^I_J}{d M}.
\end{align}
\noindent
Here, $T$ is the temperature, $m_{th}$ the threshold mass, $M$ the invariant mass and $\mu = \mu_B B + \mu_Q Q$ is the relevant chemical potential.
Moreover, $B_{J,I}$ is an effective spectral function, which is derived from the scattering phase shift $\delta^I_J$ for a given spin-isospin channel and $d_{J}=2J+1$ is the degeneracy factor for spin $J$. The yield\footnote{Since we restrict the discussion
    to pion-nucleon and pion-antinucleon interactions, the baryon and antibaryon contributions to the pressure are additive, and the corresponding yields can be uniquely identified. The thermal yield of the antibaryon channels is obtained by reversing the sign of the chemical potential $\mu$ in~\eqref{eqn:res}. Charge conjugation symmetry of the strong interaction implies that the pion-antinucleon phase shifts are identical to the corresponding $\pi N$ phase shifts. Thus, at vanishing chemical potential, the proton and antiproton yields are, as expected, equal.} given by Eq.~(\ref{eqn:res}) includes the contribution of resonances as well as that of nonresonant, correlated $\pi N$ pairs~\cite{weinhold, smat}. The spin- and isospin-weighed sum
of the empirical phase shifts from the GWU/SAID~\cite{Workman:2012hx} partial wave analysis (PWA),
and the corresponding effective spectral functions are shown in Fig.~{\ref{fig:fig1}}.

We note that resonances, like the $\Delta(1620)$ and $\Delta(1950)$  may, as a consequence of a large inelasticity, give rise to dips rather than peaks in the effective $\pi N$ spectral functions (see Fig.~\ref{fig:fig1}). In the evaluation of the proton number, there may be compensating contributions from other channels. In the next section we discuss how these can be estimated.

We recall two important aspects of applying the spectral functions shown in Fig.~\ref{fig:fig1} in thermal model studies.
First, 
at low temperatures ($T<160$ MeV), the strong Boltzmann suppression at large energies implies that thermal observables are 
sensitive to the low invariant mass part of the spectrum.
For example, at $T<160$ MeV as much as 90\%  
of the susceptibility $\chi_{BQ}$ is due to states with invariant mass $M \leq 1.6$ GeV. This underlines the importance of an accurate treatment of the low invariant mass region. By the same token, the repulsive contribution of the conspicuous dip in the effective spectral function around $M \approx 1.9$ GeV, which is associated with the $\Delta(1950)$ in the $F_{37}$ partial wave,
is almost negligible ($< 1 \%$ of the total $I=3/2$ contribution). 
By contrast, the decreasing phase shift in the $S_{31}$ channel near the $\pi N$ threshold (see, e.g., Fig. 1 of Ref.~\cite{chiBQ}) generates substantial repulsion, which reduces the $I=3/2$ contribution by about $12\%$. 

Second, in the relevant temperature range ($T\gtrsim 100$ MeV), applying the spectral functions $B_{J,I}(M)$ generally leads to a reduction~\cite{chiBQ} of the interaction strength in baryon channels,
compared to the zero-width treatment of the HRG model\footnote{As noted in Ref.~\cite{dmb}, the HRG model predictions can be accommodated in the form of Eq.~\eqref{eqn:res} by replacing
$    \delta^I_J(M) \rightarrow \pi \times \theta(M-m^{\rm res}_{JI})$ and
$    B_{JI}(M) \rightarrow 2\pi \times \delta(M-m^{\rm res}_{JI})$.}.
Consequently,
the $\pi N$ interaction contributions to the thermodynamics and to the  proton yields are reduced. 
The question is then whether the improved treatment of interactions is sufficient to remove the observed discrepancy between the HRG model and experiment. This will be addressed in the following section.

We conclude this section by noting that several of the relevant baryonic resonances have substantial branching ratios for decaying into the $\pi\pi N$ channel. Thus, one may ask how relevant the S-matrix approach restricted to elastic $\pi N$ scattering is for computing the proton multiplicity. As noted above, a consistent treatment of the $\pi\pi N$ channel in this scheme would require the knowledge of the corresponding three-body scattering amplitudes, and is beyond the scope of this letter. Nevertheless, as discussed below, we obtain an estimate of the contribution of three- and higher-body channels to the proton yield by employing LQCD results on $\chi_{BQ}$. 

\begin{figure*}[ht!]
 \includegraphics[width=0.49\textwidth]{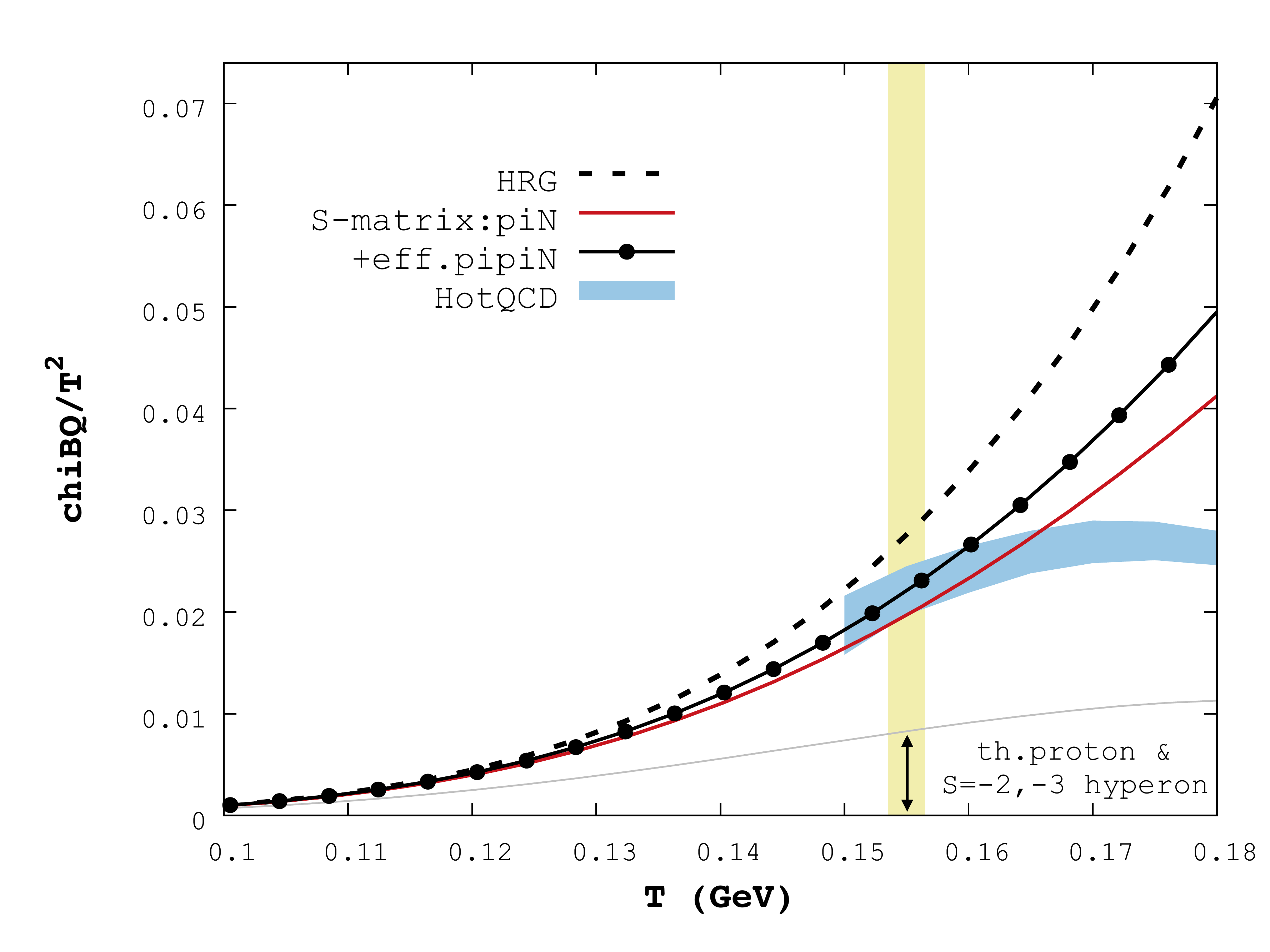}
 \includegraphics[width=0.49\textwidth]{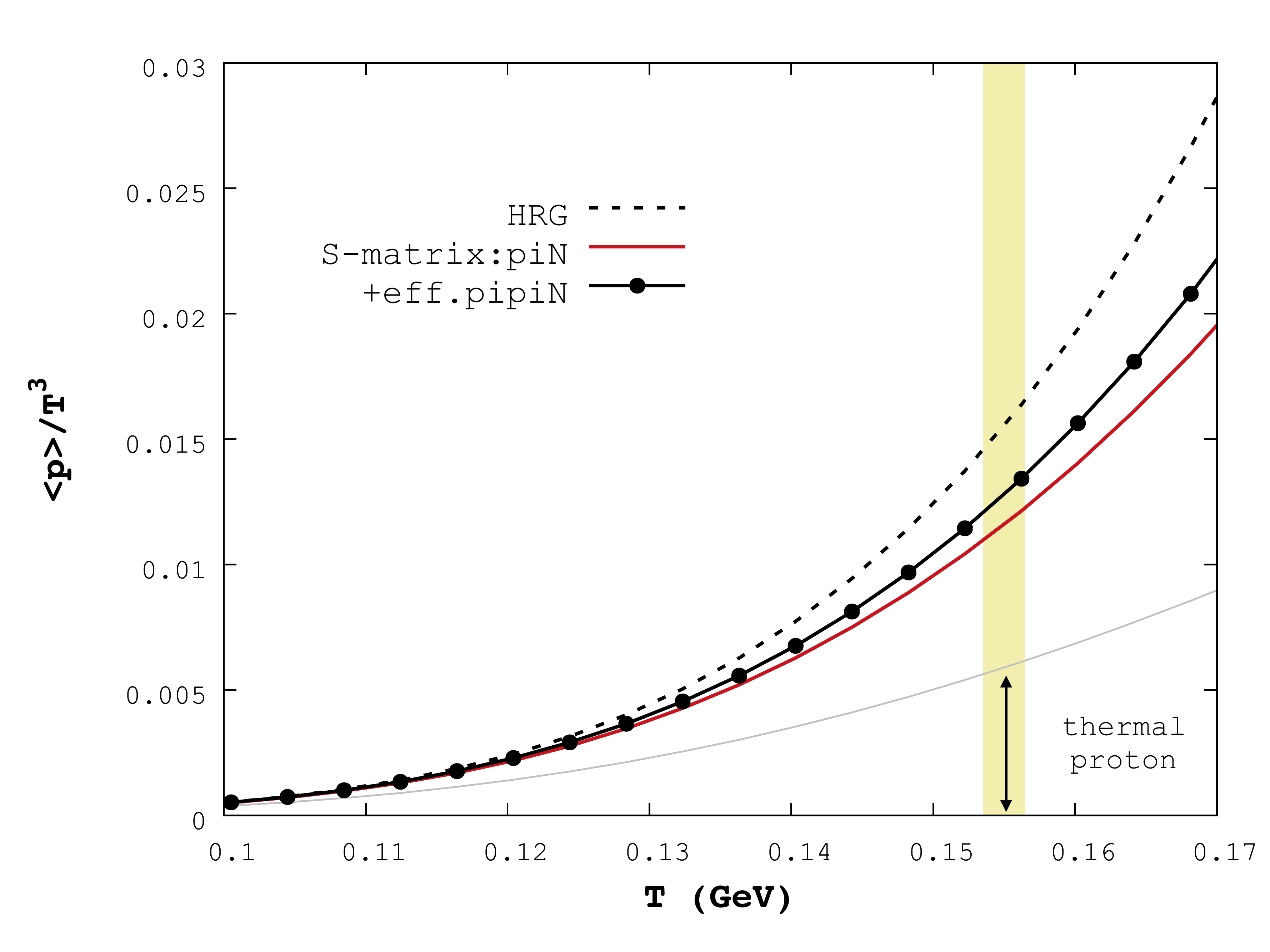}
  \caption{(color online).
  The baryon-charge susceptibility $\chi_{BQ}$~\cite{chiBQ} (left) and the proton yield (right) computed within the S-matrix formulation 
  (with and without the effective $\pi \pi N$ contribution \eqref{eqn:b3}) compared with the HRG predictions. 
  The continuum extrapolated LQCD result for $\chi_{BQ}$ is adopted from Ref.~\cite{lqcd}.
  The  vertical  band  indicates  the  location of the crossover  region. The width of the band is given by  the  uncertainty of the  pseudo-critical  temperature $T_{pc}= 156.5\pm 1.5$ MeV \cite{Steinbrecher:2018phh}.}
\label{fig:fig2}
\end{figure*}

\begin{figure*}[ht!]
 \includegraphics[width=0.49\textwidth]{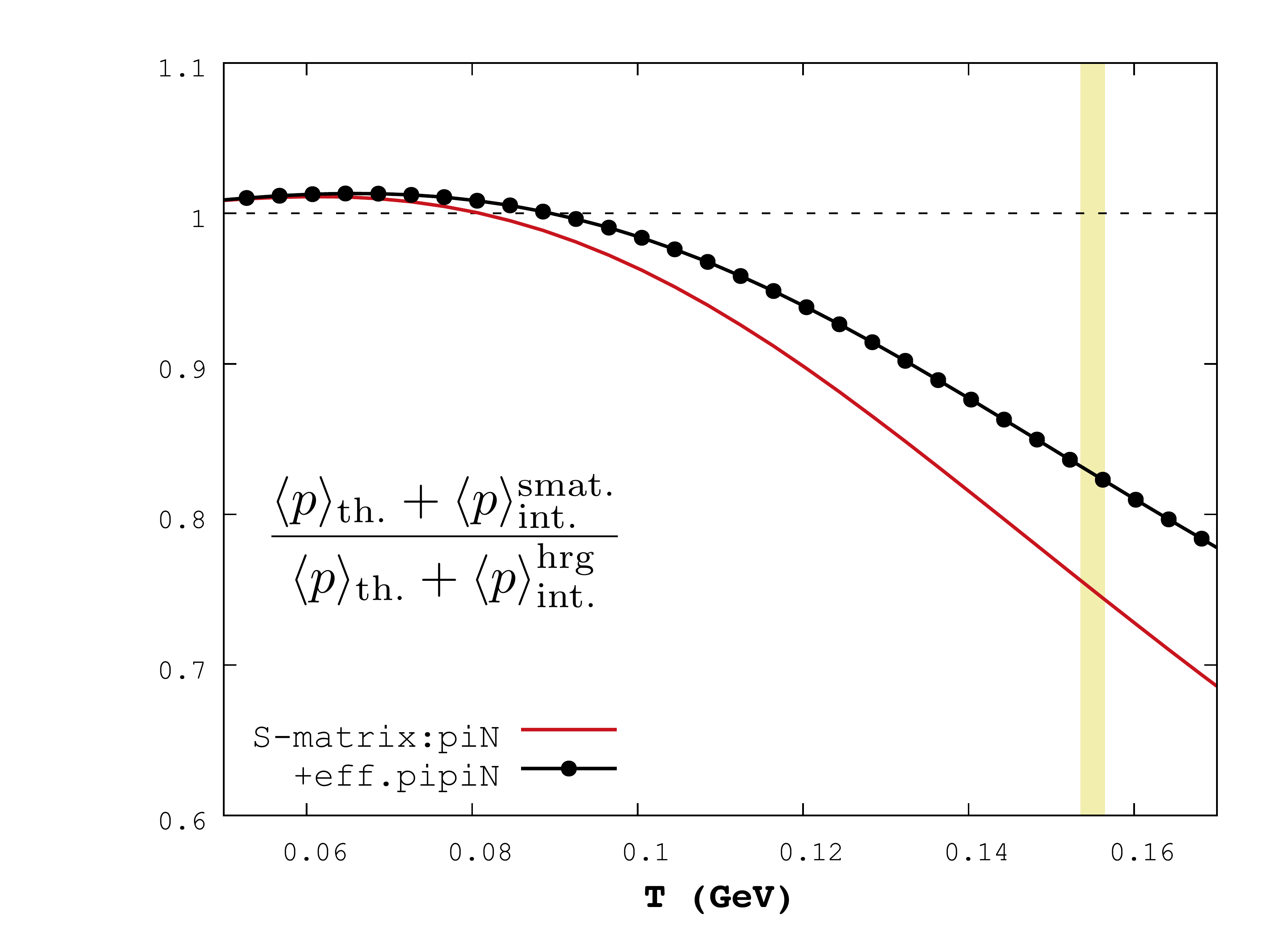}
 \includegraphics[width=0.49\textwidth]{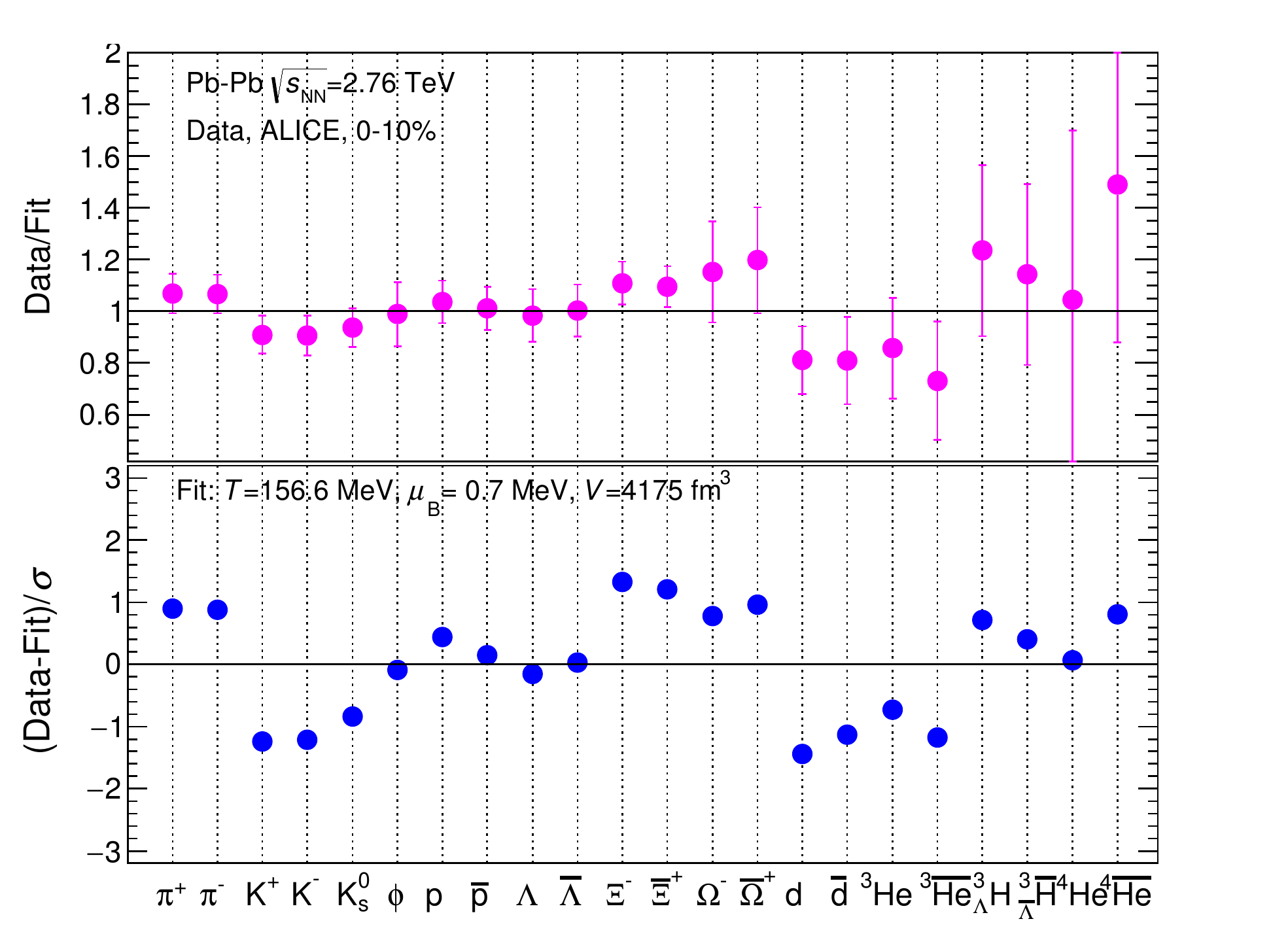}
  \caption{
    Left: The ratio of proton yields in the S-matrix and the HRG approaches, which is used as a correction factor in the statistical model. 
    Right: The ratio of data on the multiplicity per unit rapidity, dN/dy, for hadrons measured at midrapidity to the statistical hadronization results, including the  S-matrix corrections to the proton and antiproton production yields. The  lower panel shows the difference between data and model results, normalized by the standard deviation $\sigma$ of the data for a given species.
   }
\label{fig:fig3}
\end{figure*}

\section{The proton yield}

Within a statistical model, the total proton yield can, as noted above, be split into a purely thermal part and the
contribution from resonance decays\footnote{Here we use the notation ``resonance'' generically for the contribution from $\pi N$, $\pi\pi N$ {\em etc.} interaction channels with the corresponding quantum numbers.}:
\begin{align}
\label{eqn:yield}
\langle p \rangle \approx \langle p \rangle_{\rm th.} + \langle p \rangle_{N^*} + \langle p \rangle_{\Delta} + \ldots.
\end{align}
\noindent We focus on non-strange  resonances, as they provide the major part of the decay contribution. In the ALICE data~\cite{Abelev:2013vea}, the protons emanating from weak decays of hyperons are eliminated. 
The proton yield from a given resonance can be related to its thermal abundance. Consider first the case of a two-body decay into a nucleon and a pion. 
Isospin coupling determines the contribution of the various charge states
to a final state with a proton,  
i.e.,
\begin{align}
  \label{eqn:cg}
  \begin{split}
  &\langle p \rangle_{N^*} = \frac{2}{3} \langle N^*_{Q=0} \rangle + \frac{1}{3} \langle N^*_{Q=1} \rangle \\
  &\langle p \rangle_{\Delta} = \langle \Delta_{Q=2} \rangle  + \frac{2}{3} \langle \Delta_{Q=1} \rangle  + \frac{1}{3} \langle \Delta_{Q=0} \rangle.
  \end{split}
\end{align}
\noindent The thermal abundance of a resonance state is readily computed using Eq.~\eqref{eqn:res}.

A further simplification enters when we focus on the freeze-out condition relevant for describing 
hadron production in central nucleus-nucleus collisions at ultra-relativistic energies, such as those at the LHC. In this case, all freeze-out chemical potentials are practically zero, and by isospin symmetry
all charge states would acquire the same thermal abundance, leading to
\begin{align}
  \label{eqn:res_yield}
  \begin{split}
  &\langle p \rangle_{N^*} \approx \frac{1}{2} \langle N^* \rangle \\
  &\langle p \rangle_{\Delta} \approx \frac{1}{2}  \langle \Delta \rangle.
  \end{split}
\end{align}
\noindent Note that Eq.~\eqref{eqn:res_yield} remains valid for describing decays beyond the two-body case, as long as the final state contains one nucleon\footnote{The coefficients in front of each charge state in Eq.~\eqref{eqn:cg}, on the other hand, will in general be altered.}.

A comparison of (\ref{eqn:res}) with the S-matrix expression for the contribution of the spin-isospin channel $(J,I)$ to the baryon-charge susceptibility~\cite{chiBQ} at $\mu=0$, 
\begin{align}
\label{eqn:chibq0}
\begin{split}
\Delta\chi_{BQ} &= d_{J}\,\sum_{I_z,B=\pm 1} B\, Q \,\int_{m_{th}}^\infty d M\,\frac{1}{2 \pi} \, B_{J,I}(M)  \\ &\times \int \frac{d^3 p}{(2 \pi)^3}  
\, \frac{e^{(\sqrt{p^2+M^2})/T}}{(e^{(\sqrt{p^2+M^2})/T}+1)^2},
\end{split}
\end{align}
where $I_z$ is the isospin projection, shows that in the light quark sector the interaction contributions to the proton number and to $\chi_{BQ}$ are closely related. In fact, in the Boltzmann approximation, the two are proportional,
\begin{align}
  \label{eqn:chibq}
  \begin{split}
    &T \, \langle  \chi_{BQ} \rangle_{N^* + \bar{N}^*} 
    \approx 2 \, \langle p \rangle_{N^*} \\
  &T \, \langle \chi_{BQ} \rangle_{\Delta+\bar{\Delta}} 
  \approx  2 \, \langle p \rangle_{\Delta}.
  \end{split}
\end{align}
\noindent Note that at $\mu=0$, nucleons and antinucleons contribute equal amounts to $\chi_{BQ}$. Hence, the factor 2 in (\ref{eqn:chibq}). 

We can now estimate the contribution of the $\pi \pi N$ and other channels to the proton number by using (\ref{eqn:chibq}) and taking the difference of the LQCD\footnote{We note that the physical size $L$ of the lattice is large enough to provide a reliable estimate of the $\pi\pi N$ contribution to the S-matrix. In Ref.~\cite{lqcd} $LT = 4$, which implies that $L\simeq 5$ fm $\gg 2\, m_\pi^{-1}$ at $T=155$ MeV.} and S-matrix results for $\chi_{BQ}$~\cite{lqcd,chiBQ}. 
As discussed in \cite{chiBQ}, the channels with strangeness $|S|\!=\!1$ do not contribute to  the baryon-charge susceptibility at $\mu=0$. Moreover, the $|S|=2,3$ states are included in the LQCD result  and explicitly added to the S-matrix evaluation of $\chi_{BQ}$ and thus effectively eliminated in the difference of the two\footnote{Since the total contribution of the $|S|=2,3$ channels is fairly small, this approximate treatment is justified.}. Consequently, this procedure selects the three- and higher-body S-matrix contributions in the light-flavor sector. The removal of $S\!\neq\! 0$ channels is in line with the ALICE analysis, where weak decays of hyperons are excluded from the proton yield~\cite{Abelev:2013vea}.

We employ an elementary model with a structure-less interaction vertex~\cite{smat} to parametrize the temperature dependence of the three- and more-particle S-matrix. The resulting contribution to the proton number is given by 
\begin{align}
  \label{eqn:b3}
  \begin{split}
    \langle p \rangle_{\pi \pi N}  &= \int_{m_{th}}^\infty d M\, \frac{1}{2 \pi} \, B_{3+}(M) \\&\times\,\int \frac{d^3 p}{(2 \pi)^3}  \, \frac{1}{e^{(\sqrt{p^2+M^2})/T}+1},
    \end{split}
\end{align}
where $B_{3+}$ is the corresponding effective spectral function. We approximate $B_{3+}$ by the leading three-body term, with an effective interaction strength $\lambda_3$, 
\begin{equation}     
    B_3(M)  = \lambda_3 \times \frac{d}{d M} \phi_3(M^2). 
\end{equation}
\noindent Here $\phi_3(M^2)$ is the 3-body Lorentz invariant phase space of the $\pi \pi N$ system,
\begin{align}
  \begin{split}
    \phi_3(M^2)=&\int \frac{d^3k_1}{(2 \pi)^3} \frac{d^3k_2}{(2 \pi)^3} \frac{d^3k_3}{(2 \pi)^3} \frac{1}{8 E_1 E_2 E_3}  \times \\
                &(2 \pi)^4\, \delta^{(4)}(P-k_1-k_2-k_3),
  \end{split}
\end{align}
\noindent  where the  invariant mass is given by $M^2=P^2$. By fitting the S-matrix expression for $\chi_{BQ}$ to the LQCD results at $T = 156 $ MeV, we find the effective coupling $\lambda_3 = 0.955 \, {\rm MeV}^{-2}$. 

The resulting baryon-charge susceptibility and proton yields computed within the S-matrix formalism are shown in Fig.~\ref{fig:fig2}. We observe that a consistent treatment of $\pi N$ and $\pi\pi N$ interactions yields a significant reduction of the proton yield relative to the HRG result. Close to the freeze-out temperature $T_f \approx 156$ MeV, the reduction due to the elastic $\pi N$ scattering is  approximately 26\%, while the multi-pion channels increase the proton multiplicity, leading to a net decrease of about 17\%. The corresponding reduction of the pion yield is only about 1\%, and therefore neglected in the following. 

To cast the results into a form that is more convenient for the subsequent thermal model analysis, 
we construct the ratio of the yields as shown in Fig.~\ref{fig:fig3} (left). This can be interpreted as a correction factor for the light flavor baryon sector of HRG-based models. 

It was recently suggested~\cite{Vovchenko:2018fmh} that the proton anomaly may be resolved by taking the effects of resonance widths into account by means of energy-dependent Breit-Wigner parametrizations. However, this approach is incomplete, 
since, in such a scheme, nonresonant contributions are not accounted for. These include purely repulsive channels as well as background phases in resonant channels. While such interactions do not, in general, alter the hadron excitation spectrum appreciably, they can modify the thermodynamics in a nontrivial manner~\cite{weinhold, smat, exvol}. Thus, the proposition of Ref.~\cite{Vovchenko:2018fmh}, which includes only modifications of the resonant contributions, does not provide a  resolution to the proton anomaly\footnote{We stress that the thermodynamics is uniquely determined by the S-matrix~\cite{dmb}. The resonant part considered in~\cite{Vovchenko:2018fmh} constitutes only a part of the interaction contribution~\cite{weinhold}. Moreover, the split of $B_{J,I}$ into a resonant and a nonresonant part is not unique, and hence model dependent.}.

An illustrative example of the importance of background phases is provided by $\pi N$ scattering in the $S_{31}$ channel. Although the $\Delta(1620)$ is a relatively sharp resonance, the net contribution of the $\pi N$ S-matrix in this channel is repulsive. The main effect is caused by the negative background phase below the resonance. 
The S-matrix formulation of statistical mechanics naturally includes both resonant and nonresonant contributions 
by using the effective spectral function~\eqref{eqn:bfunc} to describe the effect of interactions. 

We note that there is no empirical information on background phases of the S-matrix from inelastic scattering. Consequently, a reliable determination of the contribution of $\pi\pi N$ and other channels to the proton number based on empirical scattering data is not possible. As discussed above, we bypass this problem by using the lattice results on $\chi_{BQ}$ to estimate these contributions.

We implement the correction factor to the proton yield in the statistical model, 
and perform a global fit to the hadron multiplicities measured in $\sqrt{s_{NN}} = 2.76$ TeV central Pb+Pb collisions at the LHC. In the hadron-resonance gas base-line model neither excluded volume corrections nor widths of non-strange baryon resonances are taken into account.
A comparison of the resulting hadron yields with the experimental data is shown in Fig.~\ref{fig:fig3} (right).
The current fit yields an improved $\chi^2=16.9$ (per 19 d.o.f),  compared to 29.2 obtained previously \cite{lhc1}.
At the same time, the freeze-out parameters are within errors in  agreement with previous findings, while the original 
discrepancy in the proton yield is 
eliminated. The corresponding reduction of the pion yield is very small (see above). This demonstrates the robustness and importance of the S-matrix corrections, and calls for further work along these lines for other hadrons, exploring, in particular, the remaining smaller differences seen, e.g., in the kaon and cascade yields.

 We close this section with a comment on the role of hyperon decays for the proton yield. As noted above, the weak decays of hyperons are detected in the analysis of the ALICE data~\cite{Abelev:2013vea}, and the corresponding contributions to the proton yield are subtracted. In particular, this applies to the ground-state hyperons $\Lambda, \Xi$ and $\Omega$, which are identified and listed in Fig.~\ref{fig:fig3} (right). Still, higher mass hyperons, for which strong decay channels are open, contribute to the proton number. However, in the HRG base-line model the  proton yield at freeze-out from strong $\Lambda^*$ and $\Sigma^*$ decays is only about five percent of the total yield. 

Moreover, an S-matrix treatment of the thermodynamics of the coupled ($\bar{K}N, \pi \Lambda, \pi \Sigma$) system was recently initiated~\cite{chibs}. Preliminary results indicate that the contribution of strong decays of $|S|\!=\! 1$ hyperons represent a few percent of the total proton number, in agreement with the HRG estimate. 

Consequently, we find that the contribution of strong hyperon decays is subleading to the effects discussed in this letter. Therefore, we conclude that an S-matrix treatment of strangeness channels would not modify our central result.
Nevertheless, a more exhaustive global analysis, covering the interactions of hadrons of various species, should be performed, in order to assess the influence of the present approach on other hadron yields, such as those of kaons and $\Xi$ hyperons. This will be addressed in future work.

\section{Conclusions}

In this work, we computed the contribution of pion-nucleon and multipion-nucleon interactions to the proton yield
by employing the S-matrix formulation of statistical mechanics. By implementing the essential features of the empirical $\pi N$ scattering -- the effects of broad resonances and the presence of nonresonant contributions -- and using LQCD results on the baryon-charge susceptibility in the statistical model, we find a reduction of the proton yield relative to the HRG result.
It is then in excellent agreement with experiment. 

A natural extension of this work is to employ the S-matrix scheme to
the scattering of various hadron species \cite{Klink:1998gy, Ronchen:2012eg, Mai:2017bge}.
This is an active area of research  
and the approach presented in this work provides a useful bridge for adapting these studies into the thermal model analysis of particle production in nucleus-nucleus collisions.



\section*{ Acknowledgment } 

K.R. acknowledges stimulating discussions with Frithjof Karsch. 
PML thanks Olaf Kaczmarek, Christian Schmidt and Frithjof Karsch for the warm hospitality at the Bielefeld University. BF thanks Wolfram Weise for useful discussions.
This work was supported in  part by the Extreme Matter Institute EMMI at GSI, the Polish National Science Center NCN under Maestro grant
DEC-2013/10/A/ST2/00106, and the Deutsche Forschungsgemeinschaft (DFG) through the grant CRC-TR 211 "Strong-interaction matter under extreme conditions".
This work is part of and supported by the DFG Collaborative Research Center ''SFB1225/ISOQUANT''.


\bibliographystyle{plain}

\end{document}